\documentclass[usenatbib]{mn2e}
\usepackage{graphics,amsmath}
\usepackage{rotating}
\title[The TF relation at $z\sim0.85$]{The Tully-Fisher relation of galaxies at $z\sim 0.85$ in the DEEP2 survey}
\author[K.~Chiu, S.P.~Bamford, A.~Bunker]{Kuenley Chiu$^{1,2}$, Steven P. Bamford$^3$, Andrew Bunker$^{1}$ \\
$^{1}$\,School of Physics, University of Exeter, Stocker Road, Exeter EX4 4QL\\
$^{2}$\,chiu@astro.ex.ac.uk\\
$^{3}$\,Institute of Cosmology, University of Portsmouth, Portsmouth, PO1 2EG 
}
\begin{document}

\date{Accepted 20 Feb 2007 for publication in MNRAS}

\pagerange{\pageref{firstpage}--\pageref{lastpage}} \pubyear{2006}

\maketitle

\label{firstpage}

\begin{abstract}
  
    Local and
  intermediate redshift ($z\sim0.5$) galaxy samples obey well correlated
  relations between the stellar population luminosity and maximal
  galaxy rotation that define the Tully-Fisher (TF) relation.  
  Consensus is starting to be reached on the TF
  relation at $z\sim0.5$, but work 
  at significantly higher
  redshifts is even more challenging, and has been limited by small
  galaxy sample sizes, the intrinsic scatter of galaxy properties, and
  increasing observational uncertainties.  
  We present here the TF measurements of 41 galaxies at relatively high redshift, 
   spectroscopically observed with the Keck/DEIMOS
  instrument by the DEEP2
  project, a survey which will eventually offer a large galaxy sample
  of the greatest depth and number yet achieved towards this purpose.
  The 'first-look' sample analyzed here has a redshift range of
  $0.75<z<1.3$ with $\langle z \rangle = 0.85$ and an intrinsic
  magnitude range from $M_B$ of $-22.66$ to $-20.57$ (Vega).  We find that
  compared to local fiducial samples, a brightening of $1.5$
  magnitudes is observed, and consistent with passive evolutionary models.

\end{abstract}

     
\section{Introduction}
\label{sec:intro}

The observed evolution of the Tully-Fisher relation \citep[TF;][]{tully77} 
has the potential to provide considerable insight into
how the properties of today's galaxy population were established.  In
particular, this relation between the luminosity and rotation velocity
of disk galaxies -- probed through the total mass-to-light ratio of
galaxies -- reveals links 
between dark matter haloes and the stellar
populations which form within them.  This should ideally allow us to
distinguish between the various proposed theories of galaxy formation
and evolution.

In the hierarchical structure formation paradigm of cold dark matter
(CDM), small galactic haloes ($10^{9-10}\,M_{\odot}$) first condensed
out at peaks in the near-uniform dark matter distribution. 
 Eventually, through sufficient numbers of mergers, these sub-units
accumulated to produce the massive haloes inhabited by the $\sim L*$
galaxies that dominate the stellar mass density from $z\sim1$ to the
present.  One possibility is that these first, small haloes were the sites of
significant early star formation \citep{mo98,sommer03}, 
   and the resulting stellar mass subsequently accumulated through mergers
  (along with their dark matter haloes) into more
massive galaxies without the need for substantial later
star-formation.  In this case, the growth of stellar mass in
individual galaxies was driven primarily by mergers of already
luminous sub-units, and hence the observed mass-to-light ratio should
have been relatively steady over time, evolving only slowly due to the
opposing actions of passive aging of the stellar populations and
residual star-formation.

In contrast, there is some growing evidence that star formation was
suppressed in smaller haloes at early times \citep{nelan05}.
 The suggested causes of this range from the increasing 
ultraviolet background \citep{wyithe06}, the
effects of star formation feedback \citep{springel05} (particularly from
the theoretically expected initial population of very-massive,
zero-metallicity stars), or even AGN preferentially
enhancing star formation in more massive haloes \citep{silk05}.
Whichever mechanism is responsible, this implies that galaxies must
have formed a substantial amount of their stellar mass afterwards, at later times,
following the many mergers that assembled most of their dark matter haloes.
This is somewhat reminiscent of the old
monolithic collapse model for massive galaxy formation
\citep{eggen62}, and implies substantial evolution in the
mass-to-light ratio of galaxies -- as stars form in the already
assembled massive halo.  This would also require more star formation per unit mass in smaller
haloes than in more massive ones at later times, the so called
`downsizing' effect. 

The relative order of mass assembly and star formation is cleary of
significant interest.  Therefore, examining evolution in the
mass-to-light ratio of galaxies has been pursued as an important
indicator of the galaxy formation mechanism.  Such evolution is often
probed through observed changes in fundamental scaling laws versus
redshift, such as the Faber-Jackson and Kormendy relations for
ellipticals \citep[believed to be projections of a fundamental
plane,][]{djorgovski87}.  For spiral galaxies, the Tully-Fisher relation (TF) 
\citep{tully77} is the primary scaling law.
Statistically well-populated TF diagrams over a range of redshifts
should be able to test the various galaxy formation options through
changes in the slope, intercept, and scatter of the fitted TF
relation.

Tully-Fisher relations have been relatively well characterized for
galaxies nearby, originally for distance indicator purposes
\citep{pierce92,giovanelli97,dale01}, but recently in a more
inclusive manner for assessing galaxy evolution, and to compare with
higher redshift work \citep[e.g.,][]{kannappan02}. A number of studies
have measured the TF relation at intermediate redshifts ($z\sim
0.4$--$1$) and attempted to infer its evolution with cosmic time.
However, a consensus on the TF relation evolution over the past Hubble
time (out to $z\sim 1$) has not yet been achieved, although very recent work
such as \citet{weiner06} have greatly improved upon the few existing high redshift studies.  
Difficulty arises
due to the relatively low numbers of galaxies typically used;
the intrinsic scatter in the TF relation means that many tens of
galaxies are needed per redshift bin for a reliable measurement,
particularly to constrain the slope of the relation.  The primary
cause of discrepancies between studies appears to be differing
selection criteria, and the manner in which these have been corrected
for, if at all.  The applied corrections can vary widely
depending on the percieved aim of each particular study.  In addition,
the use of different local relations, TF fitting methods, and internal
extinction corrections has complicated comparisons between studies.

As a result, various groups have identified significant evolution in
the $B$-band TF relation out to $z\sim1$ \citep[e.g.,][]{vogt96,vogt97,rix97,
simard98,barden03,bohm04,bohm06,nakamura06},
or state an upper limit \citep{bamford06} on the evolution, while
others find no evolution once selection effects have been accounted for
\citep{simard99, vogt01} or kinematically disturbed galaxies discarded
\citep{flores06}.  However, a consistent picture is begining to
emerge, as these studies become more mature.  One fairly certain
aspect is that there has been rather little evolution in the intercept
of the TF, ranging from no evolution to a modest brightening of
$\sim 1$ mag at fixed rotation velocity by $z \sim 1$.  This is much
less than would be expected if the star formation rates of massive
disk galaxies have evolved as strongly as is measured for the global
star formation rate density of the universe \citep{bamford06}.

Constraining evolution in the TF slope is more difficult, though again
any evolution is fairly modest.  Any overall brightening of the TF
that is observed appears to be mainly driven by low mass galaxies, as
noticed by \citet{bamford06}, identified as an additional population
by \citet{vogt01}, and reflected in the TF slope change measured by
\citet{bohm04,bohm06}.   This is
particularly the case when considered along with the observed lack of
evolution in the $K$-band and stellar mass TF relations found by
\citet{conselice05}, implying that while the stellar to total mass of
galaxies stays roughly constant, the colour, and hence
luminosity-weighted age or fraction of stars that have recently
formed, has evolved, possibly with a dependence on galaxy mass.
However, it is still unclear how much of this effect is actually due
to differential luminosity evolution at a given rotation velocity, or
a symptom of underestimating the rotation velocities for an increasing
population of kinematically disturbed galaxies at higher redshifts.

\begin{figure}
\includegraphics[scale=0.35,angle=0]{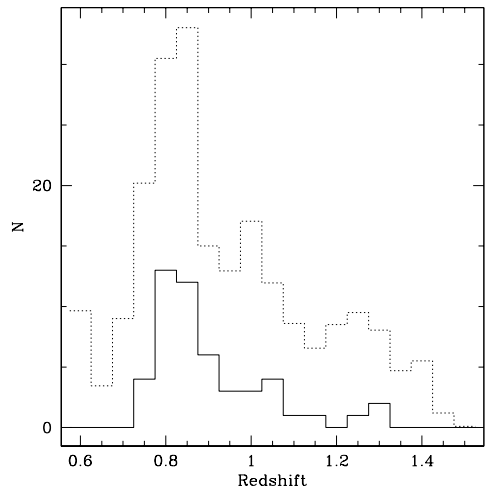}
  \caption{Redshift distribution of galaxies measured in this work (solid line).  Dotted line indicates distribution of full DEEP2 spectroscopic sample, prior to TF analysis selection, discussed in text.  The full sample distribution has been scaled by 0.05 for comparison purposes.  \label{zdistfig}}
\end{figure}

\begin{figure*}
\resizebox{0.49\textwidth}{!}{\includegraphics*[125,165][489,739]{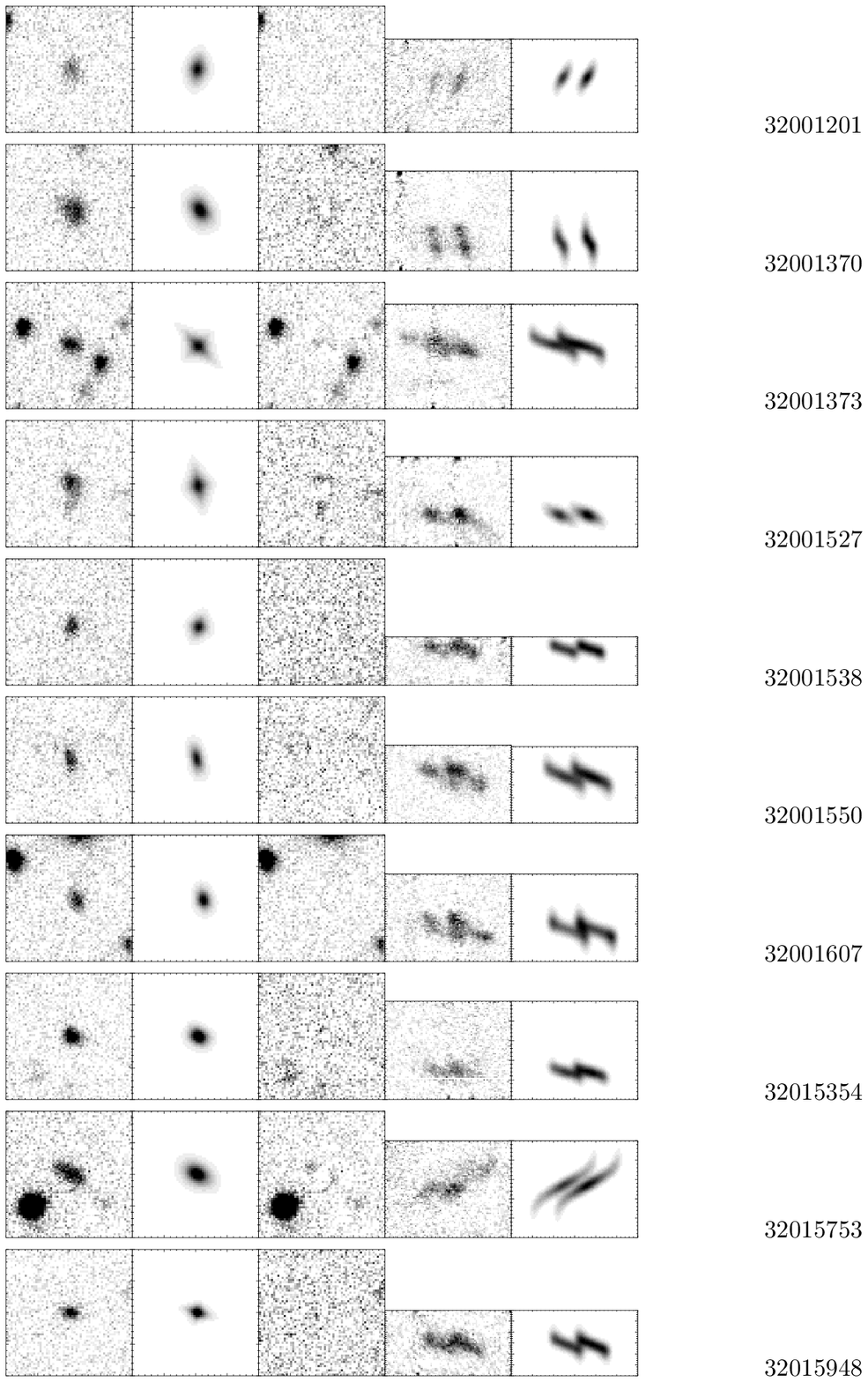}}
\resizebox{0.49\textwidth}{!}{\includegraphics*[125,165][489,739]{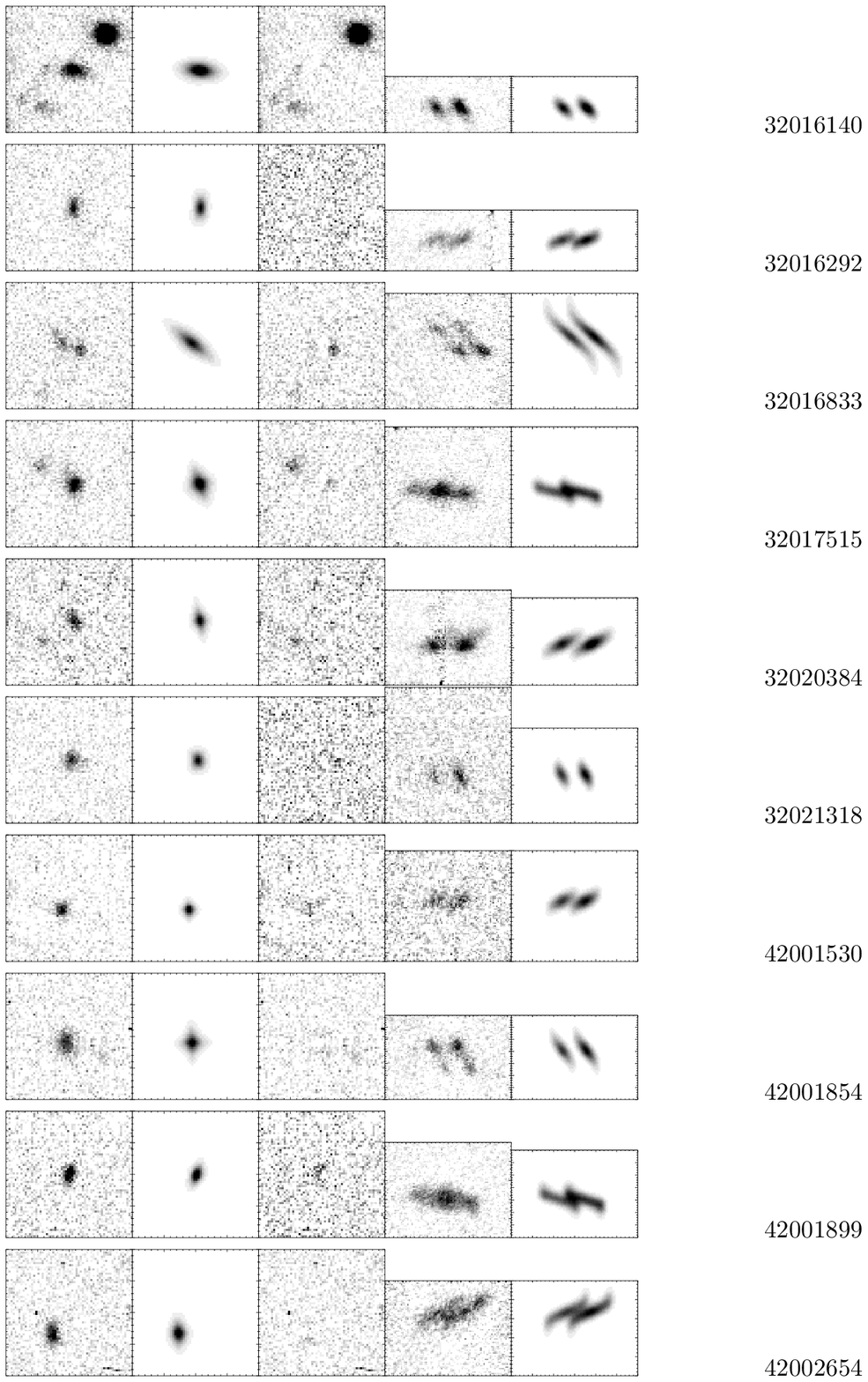}}
\caption{ Image and spectral model fitting results for
  the galaxies in the present TF sample are shown, with
  DEEP2 galaxy identification names.  The first 3 columns illustrate a)
  the original galaxy postage stamp image, b) the fitted model using
  an exponential disk profile in GALFIT, and c) the model-subtracted
  image.  The last two columns show d) the original spectral flux at
  the 3727\AA~ [OII] line, followed by e) the model generated by the
  ELFIT2PY procedure.  Galaxy images are $60\times60$ pixels
  ($12''\times12''$), while the 2d spectra are 80 pixels wide
   ($9''$, or 25\AA).  \label{galfig} }
\end{figure*}

\begin{figure*}
\resizebox{0.49\textwidth}{!}{\includegraphics*[125,135][489,763]{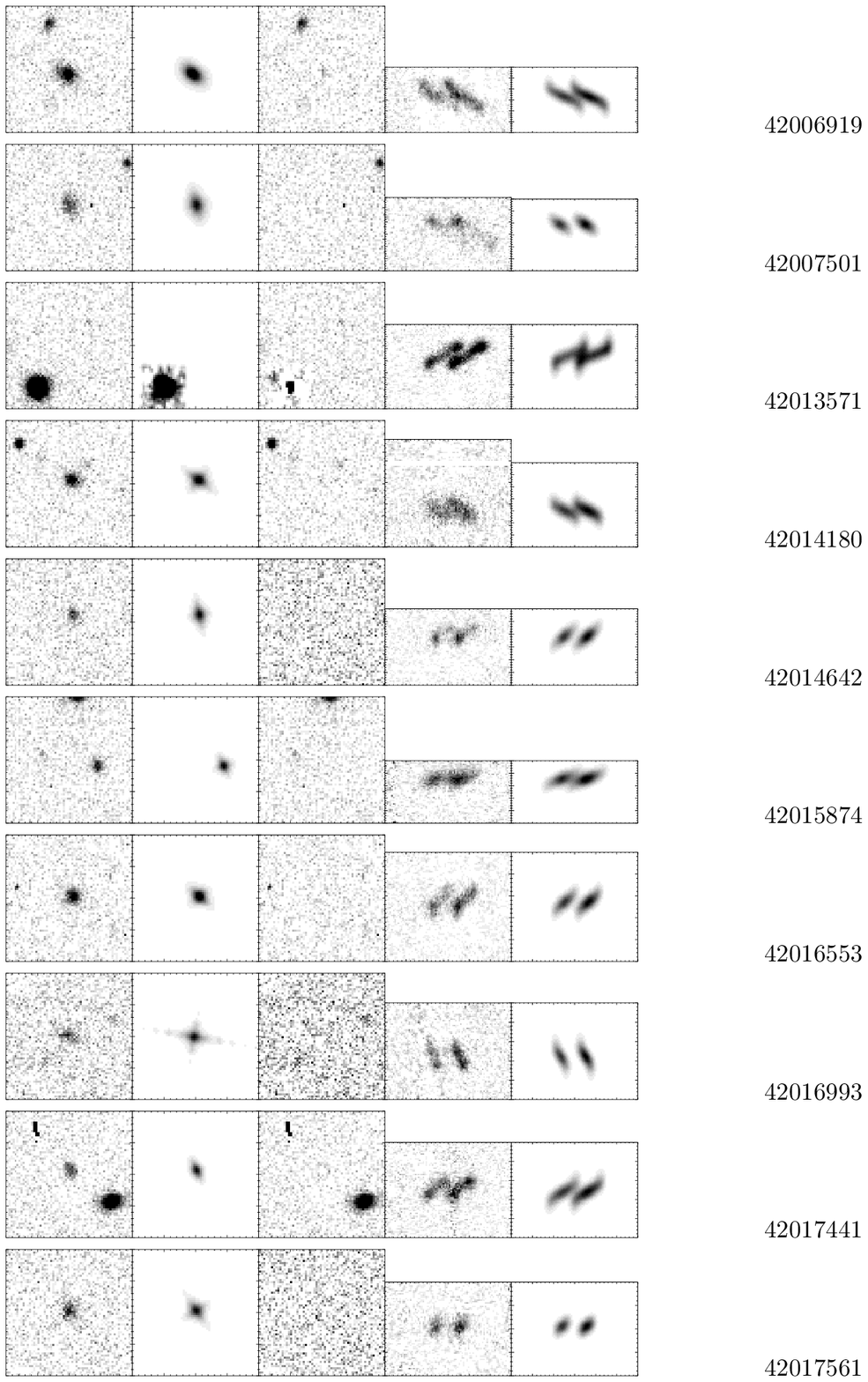}}
\resizebox{0.49\textwidth}{!}{\includegraphics*[125,135][489,763]{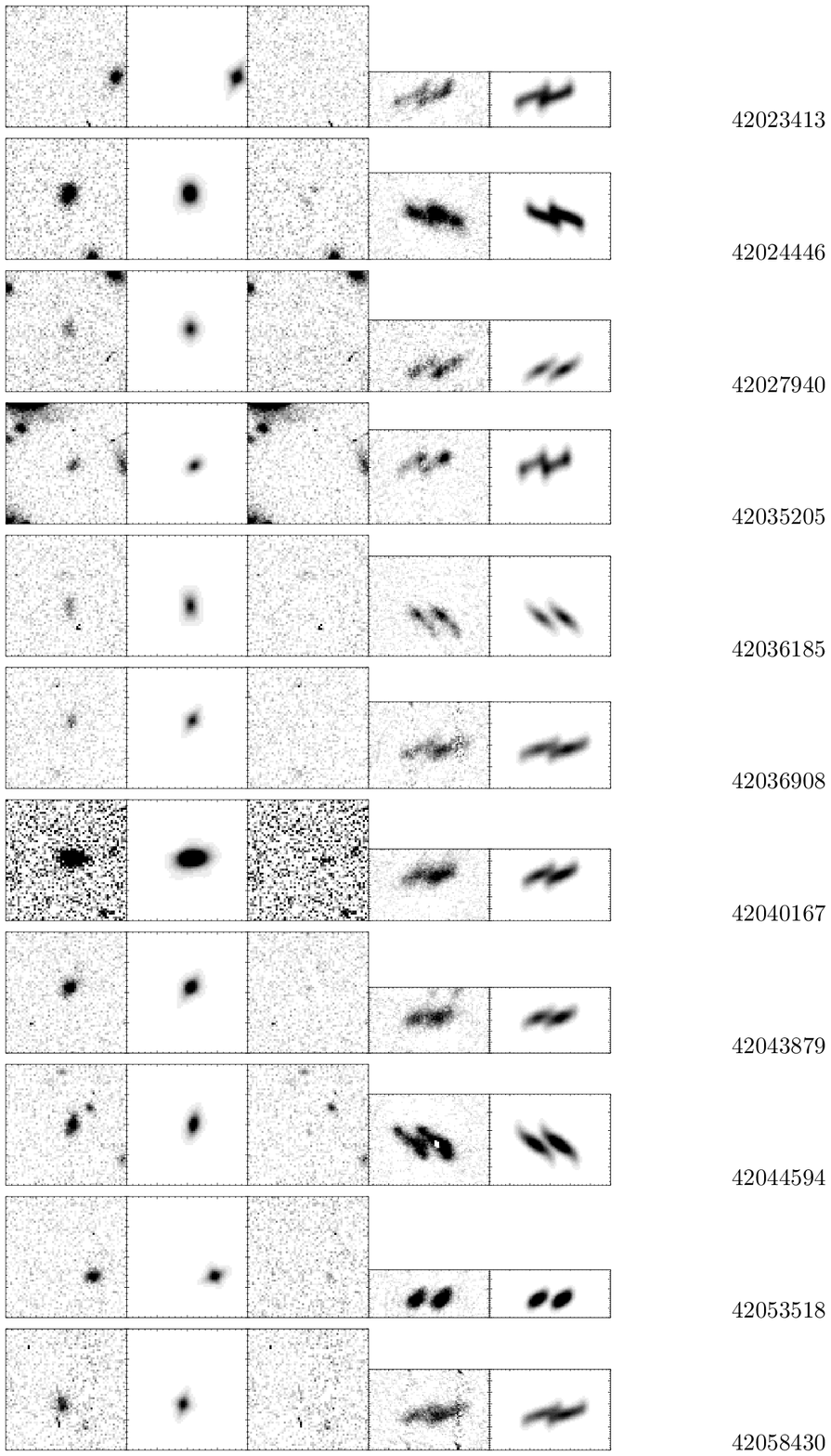}}
\caption{ -- continued.}
\end{figure*}

\citet{flores06} use integral field spectroscopy to investigate this
question, and also find that the $K$-band, or stellar mass, TF
relation has not evolved for a small sample of undisturbed disks out
to $z \sim 0.7$, while galaxies with signs of kinematical disturbances
are generally measured at low rotation velocities for their
$K$-band luminosity.  In the $B$-band, however, they show some
evidence for a brightening of $0.5$ mag, and a tentative 
change of slope consistent with the studies of \citet{bohm04}, even for
undisturbed disks.  \citet{smith04} find a consistent result from near-infrared
integral field spectroscopy for a galaxy at $z=0.82$.

A novel method in TF studies is the use of strong lensing by galaxy
clusters to magnify distant galaxies \citep{swinbank03}.  This enables
the examination of $z \sim 1$ galaxy rotation curves with resolution
comparable to $z \sim 0.1$ studies.  \citet{swinbank06} use this
technique to study six $z \sim 1$ galaxies, four of which are found to
have regular rotation curves.  These are found to define an $I$-band
TF relation very similar to that found locally, and a $B$-band TF with
similar slope to the local relation, but offset to brighter magnitudes
by $\sim 0.4$--$0.5$ mag.

While such integral field work is important in providing a complete
map of the galaxy velocity field (an improvement on the spatial
losses accompanying the usual technique of long-slit spectroscopy) the
number of galaxies observable with this method is rather limited by
present instrumental capabilities.  To measure rotation curves for a large sample
of galaxies requires spatially-resolved multi-object spectroscopy,
and currently the most efficient method is the use
of slit-masks with individual slits tilted to align with the
galaxy major axes.  The DEIMOS optical spectrograph on Keck II
\citep{faber03} provides this capability, and a major project
exploiting this instrument has been the DEEP2 project \citep{davis04}.  This
survey has undertaken spatially-resolved multi-slit spectroscopy of a large
sample of galaxies colour selected to lie at $0.7<z<1.4$.  The red
sensitivity of the instrument, coupled with its stability and lack of
fringing, means that [OII]\,3727\,\AA\ rotation curves can be traced
cleanly throughout this redshift range.

As well as spectroscopy, in order to construct a TF relation at
intermediate redshift one requires high quality imaging, to provide
magnitudes and importantly for measuring the inclination angles of the
targetted galaxies, needed to correct the rotation velocities for
projection effects.
 
In this paper, we measure the rotation curves of $41$ galaxies at
$\langle z \rangle \sim 0.85$ from the DR1 release of the DEEP2 spectroscopy, and use
them to determine the evolution of the $B$-band TF relation.  In Section~2 we
discuss the DEEP2 spectroscopy from Keck/DEIMOS, and the imaging from
CFHT. In Section~3 we detail the fitting of galaxies in
the imaging in order to derive inclinations, and the emission-line modelling to
measure the rotation velocities. Section~4 presents our resulting
Tully-Fisher relation and a comparison with previous work at low and
high redshift.  Our conclusions are in Section~5.  In this paper all magnitudes 
are expressed based on Vega normalization, and throughout we adopt
the standard `concordance' cosmology of $\Omega_M=0.3$,
$\Omega_\Lambda=0.7$, $H_0=70$ km s$^{-1}$ Mpc$^{-1}$.

\section{Data sources and processing}
\label{sec:data}

In this work, we aim to compare the fitted TF parameters for our
present sample with those for similar samples at intermediate and
lower redshifts.  The main uncertainties in constructing a TF relation
usually arise from the galaxy rotation velocities. These are
determined from the rotation curves, typically using the shapes of
emission lines measured in the spatially-resolved two-dimensional
spectroscopy. These velocities are then corrected for projection
effects due to the inclination of the galactic disk, and misalignment
of the slit from the major axis of the galaxy, both determined from
morphological fits to the imaging data.  Therefore, while the
intrinsic scatter of the galaxies themselves place limits on the
uncertainties of the fitted TF relation, galaxy imaging and
spectroscopic data of high signal-to-noise and resolution should
minimize the remaining fundamental sources of uncertainty
\citep{kannappan02}.  In particular, in this work we take advantage of
the high spatial ($0\farcs12$/pixel) and wavelength resolution
($68\,{\rm km\,s}^{-1}$) of the DEEP2 survey, combined with the wide
area imaging archive over the DEEP2 survey region from the {\it CFHT} CFH12K
mosaic imager in the $I$-band (corresponding to the rest-frame
$B$-band at $z\sim 0.9$).

\subsection{DEEP2 spectroscopy} 
 
We began our sample selection by examining the DEEP2 dataset for
suitable galaxies for TF measurement.  The DEEP2 spectroscopic
observations were based on an optical imaging survey taken from
1999--2000 using the CFH12K mosaic camera \citep{cuillandre01} in the
$BRI$ bands to a depth of $R_{AB}<24.1$.  Galaxy target selection and
slitmask design were conducted using this imaging (which covered four
fields composing a total of 3 deg$^2$),
 with color cuts in $(B-R)$ \& $(R-I)$ to optimize the selection of
galxies in the redshift range $z=0.75$--$1.4$ \citep{coil04}.

Although the DEEP2 project will eventually obtain DEIMOS spectroscopy
for some 40,000 galaxies over the total area described above, the
current data release contains $\sim 7500$ galaxies over $\sim 1.2$
deg$^2$.  Of these, approximately 4300 have confident redshifts as
measured by the DEEP2 pipeline software ($Q=4$), and we inspected
these galaxies for strong emission lines at the expected position of
{[OII]3727\AA} in each spectrum.  In this first (``1HS'') phase of the
DEEP2 project, one hour spectroscopic integrations were obtained on
each $16'\times4'$ slitmask, containing about 85 objects per mask.
Later phases of DEEP2 (``3HS'') will integrate on a smaller galaxy
sample for $>3$ hours.  After being dispersed by the 1200/mm grating,
the wavelength coverage is $6500-9100$\,\AA , so {[OII]3727\AA} is
detectable over $0.77<z<1.42$, nearly the entire redshift range
targeted by the $BRI$ colour-cuts.  The spectra have a typical
spectral resolution of $R=\lambda/\Delta\lambda_{\rm FWHM}\approx 4000
\approx 68\,{\rm km\,s}^{-1}$ determined by the $1''$ slit size,
sampled by {$\sim0.3$\AA} pixels.  The spatial pixel size is
$0\farcs119$.  Of the 4300 spectra, 388 were found to have strong
[OII] emission in the aforementioned wavelength range, and 
suitable for measuring rotation velocities using the
line fitting task with integrated single-line signal of $S/N>20$.

The spectra were provided by the DEEP2 team in the DR1 public
release\footnote{http://deep.berkeley.edu/DR1} as a fully reduced dataset,
with catalogs and wavelength-calibrated multi-extension FITS
2d-spectra and 1d spectral extractions. We did little additional
processing except that prior to the model fitting procedures described
below, we took the additional step of rectifying the 2d spectra to
have orthogonally aligned spatial and spectral axes, with constant
pixel scale.  This was carried out to correct spatial distortions 
and because the slits of the DEIMOS
instrument can be tilted to accomodate the position angles of the
selected galaxies (within $\sim30 ^\circ$, constrained by the CCD
columns and desired spatial resolution).  This results in a generally
diagonal (or even curved) wavelength solution across the spatial rows.
Using the DEEP2 wavelength solution at the given spatial center of
each galaxy in the slit mask, the solution was reproduced in all rows,
and the flux interpolated and rebinned to a constant dispersion of
$0.32$ \AA/pixel.

\subsection{CFH12K imaging}

With the sample of suitable spectra identified, we extracted images of
the galaxies for measurement of the necessary shape parameters.
Searching the CFHT archive for the three currently released DEEP2
spectroscopic fields (out of four total to be made public), we
identified useful imaging exposures taken by the CFH12K instrument.
The CFH12K camera has a platescale of $0\farcs206$/pixel, and covers a
$42'\times28'$ rectangular field of view using a $2\times6$ array of
4K$ \times $2K CCDs, for a total of $0.326$ square degrees per pointing.
Two fields, ``0230+0000'' and ``2330+0000'', had $I$-band imaging
which overlapped with most of the DEEP2 spectra where we identified
strong [OII] emission.  With a median redshift of $\langle z \rangle =0.85$ for the
galaxy sample used here (Figure \ref{zdistfig}), the choice of the
$I$-band allows the photometry to be easily converted to the $B$-band
rest-frame absolute magnitude commonly used for TF studies, with
minimal $K$-correction uncertainties.

The individual $6\times10$ minute dithered imaging exposures in each field were
combined using standard image reduction techniques, including the use
of bad pixel masks and calibration frames (``master detrend''
bias/dark/flat-field frames) identified for each camera run in the
CFH12K archive.  Cosmic ray rejection was implemented in addition to
bad pixel masking and local sky subtraction during the median
combining of the separate exposures.  Finally, each reduced image was
astrometrically corrected to the USNO-B1.0 catalog \citep{monet03},
using the WCSTools routines of D.~Mink, which automatically update
the appropriate header WCS information given a standard star catalog
and chosen reference stars in the image of interest.  This allowed
automated extraction of the image of each galaxy selected this work,
based on the J2000 RA/Dec positions given in the DEEP2 spectroscopic
catalogs.  At this point, 290 (out of 388) galaxies with strong [OII] flux
were also covered by these two CFH12K imaging mosaics, and
remained in the sample for further analysis.

\section{Data Analysis}

For the $290$ objects selected to have strong [OII] line emission
in the DEIMOS spectroscopy, and with CFH12K $I$-band images, we used
the imaging to model the morphological parameters (giving the disk
inclincation angle, $i$, and position angle), and the spectroscopy to
model the emission-line rotation curve (yielding $V_{\rm rot}\sin
i$). We describe both of these below.

The TF relation also requires absolute magnitudes ($M_B$ in this
instance), and we determined these from the $I$-band imaging,
introducing a redshift-dependent $K$-correction to match the
restframe-$B$ with the observed-$I$ filter. Finally we corrected the
absolute magnitude for intrinsic extinction in the disk galaxy,
depending on inclination and luminosity.

\subsection{Galaxy image and spectral profile fitting procedures}
\label{sec:fitting}

First, image shape modeling was carried out on each galaxy using the
GALFIT program of \citet{peng02}, in order to calculate the position
angle (PA) and axial ratio values required in subsequent steps.  For
each galaxy, a $100\times100$ pixel cutout ($12''\times 12''$) was
made from the CFH12K image, preserving astrometric and flux
information.  A point spread function (PSF) was produced by stacking
$\sim 10$ good point sources within each CFH12K array chip, resulting in a
profile with a FWHM of 2.5 pixels ($0.52''$).

GALFIT convolves this PSF with a model galaxy image based on input
initial parameter estimates ($x$, $y$ position, effective radius,
magnitude, axial ratio, position angle) and perturbs these parameters
until the residuals of the model-subtracted image are minimized.  The
initial input parameters were determined by the SExtractor program
\citep{bertin96} on the galaxy images.  An exponential disk profile was chosen as the
desired galaxy model, and the program iterated to completion,
returning each of the minimized input parameters with errors.  Of the
$290$ galaxies input to GALFIT, half returned acceptable
fits, while the remaining galaxies generally failed to converge due to
unsalvagable flux profiles (i.e., galaxies not well described by
exponential disks, or unresolved in the image).  Examples of galaxy
profile fits, and the raw/subtracted CFH12K images are shown in Figure
\ref{galfig}.

\begin{table*}  \label{galtable}
 \centering 
 \begin{minipage}{190mm}
  \caption{Extracted galaxy parameters.   Tabulated for each galaxy:
  name, coordinates in J2000, redshift, apparent magnitude,
  absolute magnitude, galaxy inclination ($i$), angular difference between
  galaxy PA and spectroscopic slit orientation, angular difference between
  slit mask and spectroscopic slit, maximal rotation velocity from systemic.     \label{galtable}}
  \begin{tabular}{@{}lccccccrrr   @{}}
  \hline
Galaxy ID & RA & Dec & Redshift & $m_I$ &    $M_B$  & $i$        & Slit--Gal   & Mask--Slit  &  $V_{\rm rot}$  \\
          & (J2000) & (J2000) & &  (Vega)&      (Vega) & $(^\circ)$ & $(^\circ)$ & $(^\circ)$ &  (km/s) \\

\hline

32001201 &   23:29:40.78  & --00:01:41.9 &    0.83580 &   $ 22.43\pm0.03$  &  $-20.85\pm0.11$    & $55.94\pm  11.1 $&    7.4	 &    28.0 &  $   79.8   \pm   7.93 $ \\ 
32001370 &   23:29:23.38  & --00:01:33.8 &    1.15568 &   $ 21.52\pm0.01$  &  $-22.66\pm0.15$    & $50.20\pm  8.90 $&   13.4	 &  --16.4 &  $   44.2   \pm   3.31 $ \\ 
32001373 &   23:29:20.31  & --00:00:58.4 &    1.03207 &   $ 22.15\pm0.01$  &  $-21.74\pm0.37$    & $55.24\pm  31.6 $&   12.5	 &  --30.0 &  $   225.3   \pm  54.07 $ \\ 
32001527 &   23:29:13.43  & --00:01:04.3 &    1.11611 &   $ 21.93\pm0.01$  &  $-22.40\pm0.14$    & $61.31\pm  6.50 $&    9.6	 &   6.8   &  $   156.3   \pm  5.94 $ \\ 
32001538 &   23:29:05.50  & --00:00:55.0 &    1.05244 &   $ 22.72\pm0.02$  &  $-20.94\pm0.17$    & $40.53\pm  25.8 $&   14.8	 &    30.0 &  $   148.3   \pm  51.18 $ \\ 
32001550 &   23:29:17.52  & --00:01:42.9 &    1.05390 &   $ 22.35\pm0.03$  &  $-21.99\pm0.13$    & $68.89\pm  4.90 $&   14.9	 &   0.0   &  $   164.0   \pm  3.14 $ \\ 
32001607 &   23:28:51.26  & --00:01:08.9 &    1.05507 &   $ 22.54\pm0.02$  &  $-21.38\pm0.13$    & $55.24\pm  9.70 $&    4.2	 &   9.8   &  $   175.1   \pm  10.86 $ \\ 
32015354 &   23:30:02.10  & +00:08:17.3  &    0.83198 &   $ 21.98\pm0.01$  &  $-21.11\pm0.09$    & $43.94\pm  13.3 $&    0.0	 &  --30.0 &  $   198.5   \pm  25.75 $ \\ 
32015753 &   23:29:50.20  & +00:06:47.1  &    0.82388 &   $ 21.71\pm0.01$  &  $-21.51\pm0.11$    & $52.41\pm  11.6 $&   --0.7   &  --30.0 &  $   361.0   \pm  29.29 $ \\ 
32015948 &   23:29:55.86  & +00:07:25.7  &    0.95589 &   $ 22.35\pm0.02$  &  $-21.50\pm0.28$    & $63.89\pm  17.9 $&  --16.1   &  --30.0 &  $   138.2   \pm  11.45 $ \\ 
32016140 &   23:29:36.71  & +00:08:56.1  &    0.80110 &   $ 21.91\pm0.01$  &  $-21.57\pm0.10$    & $65.79\pm  6.20 $&  --25.4   &  --30.0 &  $   85.8    \pm  6.53 $ \\ 
32016292 &   23:29:40.37  & +00:07:21.5  &    0.83904 &   $ 22.49\pm0.02$  &  $-21.05\pm0.15$    & $66.42\pm  10.0 $&    4.0	 &    26.7 &  $   124.9   \pm  11.75 $ \\ 
32016833 &   23:29:30.78  & +00:07:56.4  &    0.78703 &   $ 22.25\pm0.03$  &  $-21.47\pm0.18$    & $74.33\pm  7.10 $&    6.1	 &  --30.0 &  $   252.8   \pm  34.23 $ \\ 
32017515 &   23:28:57.81  & +00:08:41.9  &    0.99842 &   $ 22.02\pm0.01$  &  $-21.76\pm0.11$    & $55.24\pm  8.30 $&  --21.6   &    30.0 &  $   187.3   \pm  9.72 $ \\ 
32020384 &   23:30:19.50  & +00:11:06.8  &    1.24909 &   $ 22.63\pm0.02$  &  $-22.26\pm0.36$    & $68.89\pm  14.7 $&   29.3	 &  --13.2 &  $   170.8   \pm  12.97 $ \\ 
32021318 &   23:29:56.02  & +00:09:41.6  &    0.80140 &   $ 22.57\pm0.02$  &  $-20.57\pm0.16$    & $44.76\pm  25.2 $&   20.8	 &	 0.0 &  $    65   \pm  19.00 $ \\ 
42001530 &   02:30:43.59  & +00:22:49.2  &    0.75225 &   $ 21.89\pm0.01$  &  $-21.12\pm0.52$    & $52.41\pm  56.1 $&   24.0	 &	 0.0 &  $ 166.9   \pm  158.20 $ \\ 
42001854 &   02:30:32.79  & +00:21:53.5  &    0.80186 &   $ 21.51\pm0.01$  &  $-21.49\pm0.25$    & $42.26\pm  37.3 $&    7.0	 &    18.5 &  $   157.8   \pm  99.05 $ \\ 
42001899 &   02:30:24.87  & +00:21:36.3  &    0.90477 &   $ 21.93\pm0.01$  &  $-22.15\pm0.66$    & $72.54\pm  23.0 $&   18.2	 &    30.0 &  $   163.6   \pm  11.36 $ \\ 
42002654 &   02:30:02.52  & +00:22:49.7  &    0.80614 &   $ 21.46\pm0.01$  &  $-21.91\pm0.13$    & $60.00\pm  9.20 $&    2.6	 &    20.7 &  $   215.6   \pm  10.64 $ \\ 
42006919 &   02:30:40.80  & +00:24:27.8  &    0.89347 &   $ 21.23\pm0.01$  &  $-22.33\pm0.10$    & $55.94\pm  8.30 $&    7.1	 &  --30.0 &  $   182.4   \pm  9.12 $ \\ 
42007501 &   02:30:30.40  & +00:24:45.1  &    0.92616 &   $ 21.67\pm0.01$  &  $-22.01\pm0.12$    & $58.66\pm  9.40 $&   --4.4   &    18.3 &  $   109.1   \pm  6.92 $ \\ 
42013571 &   02:30:50.69  & +00:29:17.0  &    0.82416 &   $ 22.10\pm0.01$  &  $-20.96\pm0.07$    & $43.11\pm  10.1 $&   12.2	 &	 3.4 &  $   214.6   \pm20.95 $ \\ 
42014180 &   02:30:34.44  & +00:28:07.2  &    0.75354 &   $ 21.74\pm0.01$  &  $-21.22\pm0.17$    & $49.45\pm  19.9 $&  --25.6   &	 0.0 &  $   161.3   \pm26.94 $ \\ 
42014642 &   02:30:27.62  & +00:29:19.9  &    0.75022 &   $ 22.50\pm0.02$  &  $-20.74\pm0.62$    & $65.79\pm  36.1 $&   16.1	 &	 0.0 &  $   150.3   \pm26.90 $ \\ 
42015874 &   02:30:02.02  & +00:29:18.1  &    0.86724 &   $ 22.52\pm0.02$  &  $-20.81\pm0.50$    & $54.54\pm  51.0 $&  --14.6   &   17.4  &  $   218.9   \pm  133.10 $ \\ 
42016553 &   02:29:48.27  & +00:29:39.6  &    0.80765 &   $ 21.59\pm0.01$  &  $-21.56\pm0.25$    & $50.20\pm  27.6 $&  --18.9   &    0.0  &  $   139.4   \pm  34.85 $ \\ 
42016993 &   02:29:32.82  & +00:29:31.9  &    0.88262 &   $ 22.41\pm0.02$  &  $-21.17\pm0.31$    & $62.61\pm  22.1 $&  --23.0   &  --30.0 &  $   64.2   \pm   19.36 $ \\ 
42017441 &   02:29:17.55  & +00:28:27.9  &    0.85701 &   $ 22.49\pm0.02$  &  $-21.32\pm0.46$    & $72.54\pm  19.3 $&   --0.8   &    4.2  &  $   138.0   \pm  14.36 $ \\ 
42017561 &   02:29:13.87  & +00:27:12.9  &    0.85383 &   $ 22.26\pm0.02$  &  $-21.01\pm0.40$    & $52.41\pm  43.6 $&   --6.1   &    0.0  &  $   69.4   \pm   33.44 $ \\ 
42023413 &   02:29:59.75  & +00:33:20.1  &    0.81719 &   $ 21.46\pm0.01$  &  $-21.89\pm0.24$    & $57.99\pm  19.1 $&   20.1	 &    30.0 &  $   179.9   \pm  20.20 $ \\ 
42024446 &   02:29:28.61  & +00:32:01.6  &    0.81243 &   $ 21.04\pm0.00$  &  $-22.06\pm0.09$    & $44.76\pm  11.4 $&   --3.6   &    30.0 &  $   210.1   \pm  22.09 $ \\ 
42027940 &   02:31:01.50  & +00:35:10.5  &    0.86831 &   $ 22.39\pm0.02$  &  $-21.46\pm0.17$    & $45.57\pm  24.8 $&   --2.4   &    30.0 &  $   216.6   \pm  55.86 $ \\ 
42035205 &   02:31:09.36  & +00:40:21.8  &    0.86450 &   $ 22.76\pm0.02$  &  $-20.58\pm0.17$    & $55.94\pm  18.1 $&   14.4	 &	 7.7 &  $   134.7   \pm15.69 $ \\ 
42036185 &   02:30:59.17  & +00:40:03.4  &    0.79288 &   $ 22.53\pm0.02$  &  $-20.82\pm0.48$    & $59.33\pm  40.4 $&   25.3	 &  --30.0 &  $   203.1   \pm  59.06 $ \\ 
42036908 &   02:30:45.18  & +00:38:43.5  &    1.29168 &   $ 22.90\pm0.02$  &  $-21.84\pm0.32$    & $62.61\pm  16.8 $&    1.1	 &	 0.0 &  $   164.9   \pm13.84 $ \\ 
42040167 &   02:29:28.90  & +00:37:14.8  &    0.78000 &   $ 21.26\pm0.01$  &  $-22.09\pm0.16$    & $61.31\pm  10.4 $&   26.5	 &    30.0 &  $   171.1   \pm  8.80 $ \\ 
42043879 &   02:30:48.83  & +00:43:05.4  &    0.77335 &   $ 21.46\pm0.00$  &  $-21.71\pm0.15$    & $55.94\pm  12.5 $&   --6.0   &    11.1 &  $   184.1   \pm  14.67 $ \\ 
42044594 &   02:30:37.44  & +00:42:10.3  &    0.90493 &   $ 21.50\pm0.00$  &  $-22.48\pm0.15$    & $68.89\pm  7.30 $&   --1.5   &  --12.2 &  $   181.8   \pm  4.58 $ \\ 
42053518 &   02:30:07.45  & +00:44:46.2  &    0.92744 &   $ 21.72\pm0.01$  &  $-21.74\pm0.22$    & $48.70\pm  25.0 $&    9.8	 &    30.0 &  $   102.8   \pm  23.77 $ \\ 
42058430 &   02:30:57.37  & +00:48:03.1  &    0.97522 &   $ 21.83\pm0.01$  &  $-22.05\pm0.52$    & $60.65\pm  33.8 $&   17.0	 &  --30.0 &  $   186.5   \pm  38.66 $ \\ 
\hline									      	 	     	 									   
\end{tabular}								      	 	     	 								    
\end{minipage}								      	 	     	 								    
\end{table*}

The fitting of faint emission lines in order to extract reliable
rotation velocities is a challenging task, and highlights the
desirability of high resolution and high signal-to-noise 2-dimensional
spectra in order to determine confident galaxy rotation curves.  In
order to extract as much information as possible from the emission
lines, and to consistently model the effects of inclination, seeing,
finite slit width and instrumental resolution, we employed an enhanced
version of the ELFIT2PY routine developed by SPB \citep{bamford05},
which is based partly on the synthetic emission line fitting software
of \citet{simard98}.

The software is given the galaxy inclination, seeing, instrumental
spectral resolution, slit position angle misalignment with respect to
the galaxy major axis, slit tilt in the mask, and produces synthetic
emission lines by constructing model 2D galaxy flux and velocity
profiles, assuming a set of parameters, and then mimicking the
observations.  ELFIT2PY uses the Metropolis searching algorithm
 to find the set of parameters which minimize the
residuals between the synthetic and observed emission lines.  The
searched parameters are the rotation velocity, the emission flux
scalelength (assuming an exponential disk), the turnover radius of the
rotation curve, and the spatial and wavelength centres of the emission
line.  The use of more complicated flux profiles was attempted, but it
was found that these generally could not be constrained.  An intrinsic rotation
curve of the form $V_{\rm rot}/(r^a + r_t^a)^{1/a}$ is assumed
\citep{courteau97}, with $a = 5$ (found to best reproduce the observed
rotation curve shapes, though ill-constrained, and in agreement with
\citealt{bohm04}), and where $r$ is the radius and $r_t$ is the turnover
radius.  Such an intrinsic rotation curve is necessary in this case,
rather than the discontinuous form used previously with ELFIT2PY, in
order to adquately fit the high resolution of the
DEEP2 spectra.  Note that the emission flux scalelength and the
turnover radius of the intrinsic rotation curve were allowed to vary
independently, which was frequently found to be necessary.  The total
flux in the line was measured by integrating the background- and
continuum-subtracted image of the line, and held fixed for the fit,
with a constant [OII] doublet ratio of $f([{\rm OII} ] ~3727.092{\rm~ \AA})/f([{\rm OII}] ~3729.875~{\rm \AA}) =0.8$.  The background level
was also fixed at zero.

In previous versions of ELFIT2PY, the chi-squared comparison between
the synthetic and observed lines was done on a pixel by pixel basis in
the spectra.  However, some other groups \citep[e.g.,][]{bohm04} first
trace the synthetic and observed emission lines, by fitting the
position of the line in each spatial row independently, and then
compare these traces to produce a chi-squared value used to judge the
goodness of fit.  This has the advantage  of removing one level of
dependence on the assumed galaxy flux profile, though it still of
course plays a role in the construction of the synthetic line.  It also
weights the fit to preferentially reproduce the observed rotation
curve rather than the flux profile, which may be affected by discrete
star-formation regions and asymmetries.  However, this is at the cost
of leaving the emission scalelength unconstrained, which must
therefore by fixed relative to the rotation curve turnover radius.  
Therefore a hybrid approch was adopted to take advantage of the 
high resolution of the DEEP2 spectra (which
allow a high quality trace of the emission line to be measured), while
somewhat reducing the influence of asymmetries in the observed flux
profile which are apparently more prevalent at the high redshifts of
the current sample. 

This hybrid method combined the chi-squared values from pixel-to-pixel
comparisons of the synthetic and observed spectra, and those from
comparisons of both the trace centres and FWHM of the synthetic and
observed lines.  The variation of the observed FWHM in each spatial
row helps constrain the rotation curve as it is a measure of the
gradient of the rotation curve at each spatial position.  If the
rotation velocity is increasing quickly with radius, the observed FWHM
in that spatial row is higher.  This combined chi-squared value was
the quantity used by the Metropolis algorithm to determine the
best-fitting parameters and their confidence intervals.

The traces were performed by fitting a double Gaussian function to
each spatial row independently.  The Gaussians had a fixed separation
appropriate for the [OII] doublet at the redshift of the galaxy, and
the same FWHM.  Only the centre of the doublet, single line FWHM and
 amplitudes of the two components were allowed to vary.  Fits were
judged to be reliable if the Levenberg-Marquardt least-squares
minimisation converged from the appropriately chosen initial parameters,
if the errors on the best-fitting centre and amplitudes were sensible,
and if the best-fitting FWHM was larger than the spectral resolution.

Note that the velocity convention in this work is to use the $V_{\rm
  rot}$ ``half-velocity'' term, which specifies the maximum absolute
rotation velocity, of the asymptotic assumed intrinsic rotation curve,
with respect to the center of the galaxy, rather than the
velocity-width, $2V_{\rm rot}$, traditionally used to express the
total velocity gradient measured across the entire galaxy.  Examples
of synthetic emission-line images, created using the best-fit
parameters, appear in Figure \ref{galfig}.

The emission lines of the 290 galaxies selected thus far were
inspected visually in more detail before fitting, and those with no
sign of being spatially extended were rejected. Of the remaining galaxies, 
those with unsuccessful GALFIT
measurements of their inclination and position angle were also
rejected.  Then, only galaxies with inclinations more than $30\degr$ from
face-on, and hence a reasonable component of the rotation velocity
along the line-of-sight, were fit with ELFIT2PY.  
In addition, galaxies with a large misalignment angle 
($>30\degr$) between their major axes and the slit were rejected, 
to avoid large and uncertain corrections for this effect.
$78$ galaxies were thus remaining to be fit with ELFIT2PY.

In a number of cases, the best fitting parameters from ELFIT2PY
indicated that the rotation velocity was unreliable, for example due
to the emission flux profile being consistent with a point source, or
the end of the reliably measurable emission line occuring at smaller
radii than the best-fitting intrinsic rotation curve turnover radius.
In order to remove such uncertain galaxies from the sample we imposed
the following quality criteria.

If the scalelength of the fitted emission flux exponential disk
profile is too small, then the fit is unlikely to be useful, and thus 
is rejected.  The limit adopted is $0.14\arcsec$ -- the angle subtended
by one pixel for a galaxy observed with a slit tilt of $30\degr$, the
worst case we admit.  This corresponds to $1$--$1.2$ kpc for the
redshifts of our objects.  In order to also reject galaxies with
uncertain emission scalelengths which are close this limit, we reject
any galaxy for which the fit scalelength is consistent with being
below $0.14\arcsec$ at the $1\sigma$ level.

In order to consider how far out in each galaxy we measure the
emission line to, we define the \emph{extent} of the line as half of
the spatial distance over which the line is reliably traced, after
removing single `good' points with no `good' neighbours as spurious.
If the extent is less than that one would expect for a point source given
the seeing, which we take as $0.5''$ then we reject that galaxy from
further consideration.

Finally, we also consider whether the transition from the inner
rotation curve slope to the flat region is observed with sufficient
signal-to-noise.  This is done by requiring that the radius at which
the fit intrinsic rotation curve turns over is within the reliable
extent of the line. Furthermore, in the case where the turnover is not
observed, the turnover radius will be highly uncertain.  To exclude
such galaxies we require that the turnover radius is inconsistent with
being beyond the line extent with at least $1\sigma$ confidence.

Following the application of these quality criteria, $41$ galaxies
remained with reliable rotation velocity measurements.

\subsection{Galaxy photometry, $K$-correction, and extinction correction}

In addition to the rotation velocity, the second input required for
the TF relation is the galaxy absolute magnitude. In order to compare
with other samples, while avoiding large $K$-corrections to the 
photometry, we use the absolute magnitude in the
rest-frame $B$-band, $M_B$, in the AB zero-point system. For our
galaxies, with median a redshift of $\langle z \rangle =0.85$, this roughly corresponds
to the observed-frame $I$ filter (centered at 8100 \AA).  For each
galaxy of interest, the $I$-band photometry was extracted from the
DEEP2 photometric catalogs accompanying the spectroscopic dataset \citep{coil04}.
These catalogs were produced using the IMCAT imaging and photometric
reduction software of \citet{kaiser95}, having been finally calibrated
to the SDSS standard system. As an independent check on the $I$-band
photometry, we also verified magnitudes during our SExtractor
and GALFIT proceedures.

\begin{table*}  \label{fitstable}
 \centering 
 
  \caption{ Best-fit TF values for the sample of 41 DEEP2 galaxies. Parameters below fit the relation 
  $M_B= a + b(\log V_{\rm rot}+c) $, discussed in text, where $c$ is an additional offset factor to account 
  for the use of $\log V_{\rm rot}$ versus $\log 2 V_{\rm rot}$ in  TP00 or V01.  }
  \begin{tabular}{@{}lccccc   @{}}
  \hline
&  $a$  &  $b$   & $c$ & total scatter  & intrinsic scatter \\
&     &      &   & mags (dex)    &  mags (dex) \\
  \hline
Free-fit &$-21.57 \pm 11.67 $  & $-12.06 \pm 6.51 $ & $-2.2$ &  $1.72~ (0.143 )  $ & $1.52~ (0.126 $)\\
TP00     &$-21.58 \pm 0.17  $  & $-7.27 \pm 0.00  $ & $-2.2$ &  $1.07~ (0.147 )  $ & $0.91~ (0.125 )$\\
V01      &$-21.57 \pm 0.20  $  & $-9.00 \pm 0.00  $ & $-2.2$ &  $1.30~ (0.144 )  $ & $1.12~ (0.124 )$\\
\hline		        										       \end{tabular}												       \end{table*}

Using synthetic
template spectra of Sbc, Scd, and Im galaxy types \citep{bruzual03}, appropriate for the range of
 disk galaxies with star formation and hence [OII] emission, we
simulated flux through the rest-frame $B(z=0)$ and redshifted $I(z)$ filters.
We applied the average resulting
$K$-correction and error to each galaxy according to its redshift.  
 
Finally, we applied an inclination-dependent extinction correction to
each galaxy.  Historically, the corrections of \citet{tully85} and
\citet{tully98} have been used most commonly, the latter incorporating
an additional dependence on galaxy luminosity (or rotation
velocity).  We adopt the \citet{tully98} method, for comparison of our
work to recent TF samples at similar and lower redshifts. The
extinction is defined in \citet{tully98} to be: 
\[A = \gamma_B \log (a/b), \]
with
\[\gamma_B = -0.35 (15.31 + M_B ),\]
 where $a/b$ is the galaxy major-to-minor axial ratio.
 Table \ref{galtable} displays all of the  
 measured parameters of the galaxies in our
sample.

\section{The Tully-Fisher Relation}
\label{sec:TFdiscuss}

From the original catalog of $4300$ galaxies with high quality redshifts
in the DEEP2 DR1 data release, the aforementioned steps have reduced
the sample significantly, specifically through inspection of the 2d
spectra for [OII] emission ($N=400$), imaging data overlap matching
($N=290$), galaxy image profile modeling ($N=150$), and suitable
inclinations and slit misalignments for emission line modeling ($N=78$).
Following rejection of those galaxies for which we judge the measured
rotation velocity to be unreliable, as described in section
\ref{sec:fitting}, we now have a useful sample of $41$ galaxies with
full photometry, morphology and rotation information.  The sample 
spans a redshift range of $0.75<z<1.3$ with $\langle z \rangle = 0.85$ and an intrinsic
  $M_B$ magnitude range of $-22.66$ to $-20.57$.

\begin{figure}
\includegraphics[scale=0.5,angle=-90]{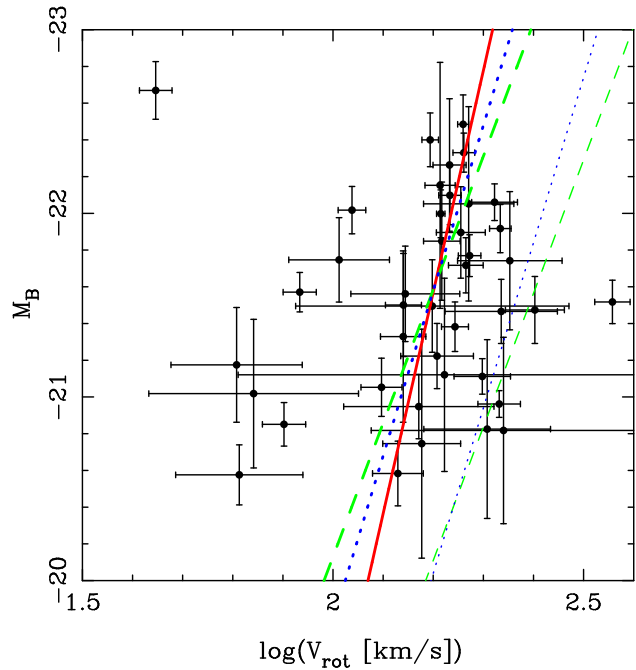}

  \caption{
  The TF relation of the 41 galaxies in the present sample are plotted
  in absolute $M_B$ magnitude versus maximal rotation velocity, and
  overlaid with the resulting linear TF fits discussed in text (in order of
  decreasing steepness, solid red line: free fit, dotted blue line: slope of
  V01, dashed green line: slope of TP00). 
  Light dotted blue and dashed green lines indicate 
  the original local V01 and TP00 relations, again in order of decreasing
  steepness.  
  Object 32001370 (extreme upper left) was excluded 
  from fits as a $>3\sigma$ outlier.  \label{newtfplot}}
\end{figure}

From this sample, we carried out several least-squares fittings
to a straight line function: $M_B= a + b\log V_{rot} $, where $a$ and
$b$ are the intercept and slope of the TF relation, respectively.  
 The slope and/or intercept value, and scatter width were minimized
as the free parameters, and as in \citet{bamford06}, we use weights of
$1/\sigma^2$ for each point, where $\sigma^2 = \sigma_{\log V_{rot}}^2
+ b^2 \sigma_{M_B}^2 + \sigma_{int}^2$, and $\sigma_{int}$ is the
intrinsic scatter of the TF relation.  One data point was rejected by sigma-clipping, 
specifically object 32001370, due to its unusual position more than $3\sigma$ from
the main sample.  Three fits were then carried out -- first, 
a free-fit of all parameters was allowed.   Because this yielded a 
relatively poorly constrained slope, two subsequent fits were made with 
slope fixed to local reference samples, specifically those of \citet{tully00} or \citet{verheijen01}.  
Each of these comparison samples employed a version of the \citet{tully98} extinction correction, 
as is used for our sample here.  
Table 2 lists these fitted parameters, and Figure \ref{newtfplot} displays the 
resulting TF distribution and fits.

These fitted parameters present several interesting aspects in comparison to the
local fiducial TF relation.  First, an overall brightening of $1.5\pm 0.2$ mag 
is seen in this sample at $\langle z \rangle \sim0.8$ compared to $z\sim0$ galaxies at present.  
This brightening is consistent with \citet{bohm04} who find an offset of $-(1.22\pm0.56~ {\rm mag})z$, as 
well as \citet{bamford06}, who find an offset of $ (-1.0\pm0.5~ {\rm mag})z$.  
The strength of the effect may be explained in part by a colour selection effect
in the DEEP2 galaxy sample, which leads to observation of relatively younger stellar population galaxies 
(and in turn, decreased mass-to-light ratios).  A larger sample from the full 
DEEP2 data may allow this colour-TF relation to be evaluated, as in \citet{kannappan04}.  
Also of interest is that
 the brightening of this sample
compared to the local fiducial samples seems to be strongest at the bright end of the 
TF relation ($M_B<-20.5$).  This is in contrast with the result of \citet{bohm04}, 
in which little brightening ($\sim0.5$ mag) is found the relatively bright galaxies, 
but significant brightening found in the faint end galaxies ($\sim1.5$ mag, $M_B\sim-18$).

 Does this magnitude offset
accord with the hierarchical picture, in which small galaxies with
pre-formed stellar populations merge into progressively more massive
systems, whose mass-to-light ratios change only slowly with the
dimming of the passively evolving galaxies?  We have calculated the possible
contribution of such passive evolution to observed changes in high
redshift galaxy luminosities and find that such dimming over time is 
consistent.  Using Bruzual \& Charlot exponential decay  
models, we find that young
galaxies at $z=0.85$ (with luminosity-weighted ages of 3--5 Gyr)
will have dimmed by $\Delta M_B \sim1.0$ mag over the 6.5 Gyr to $z=0$.  
Hence we conclude that the observed
brightening of the TF relation in this sample is consistent with pure
luminosity evolution.

We additionally find that the amount of the brightening offset is within the intrinsic
scatter of the sample, and comparable to the scatter found in the
local reference populations discussed above.  Because intrinsic
scatter in the TF relation reflects variations in the mass-to-light
ratio of individual galaxies, stellar mass fraction, and deviations
from ordered rotation, these effects may be useful to separate in
larger similar samples.

A few points are worth consideration in the future.  First, it
will be useful to examine whether TF samples are really dominated by
disk galaxies, and the extent to which unidentified irregulars may be
influencing and scattering the relation.  This may be partially influenced
by the selection of the present sample, which is based on the color cut producing the 
DEEP2 galaxies as well as the use of relatively bright galaxies in order
to accomplish the spectral and spatial fitting tasks here.  Also, while this and other
TF studies examine the blue, star-formation dominated region of the
galaxy spectrum, the redder portion of galaxy flux displays tighter
correlation and may be more useful for detection of parameter
evolution.  The effects of momentarily bright star formation 
episodes (mostly influencing the restframe blue bands) could be mitigated.  
Such work (at $z=1$ for example) would benefit from $H$ or $K$-band
imaging, corresponding to $I$ band in the restframe, as has been demonstrated
by \citet{conselice05}.  Finally, with future 
improvements in spectroscopic and imaging spatial
resolution (such as with {\it HST}), it may be possible to more thoroughly separate the disk
and bulge flux components of examined galaxies, and even address 
non-uniform star-forming mass regions.

\section{Conclusions}
\label{sec:conclusions}
 
The population of 41 reliable galaxies examined here is a significant
addition to the existing sample at $\langle z \rangle \sim 0.85$, with spectral
resolution of $R\sim4000$ allowing the determination of the TF parameters
and their evolution.  Our detection of a 1.5 mag brightening offset in
the TF intercept compared to the present $z=0$ relation is consistent with 
the previous work of \citet{bohm04}, \citet{bamford06}, and the recent results 
of \citet{weiner06}.  We suggest that this is consistent with
passive evolution of galactic sub-units that have already undergone
star-formation prior to having assembled into the large galaxies we
see today.  This is in contrast to the scenario where significant new
star formation is proposed to occur within the dark halos that
assembled first.

Even if galaxies experience only short episodes of star formation
activity, and then evolve passively without further mergers, we have
demonstrated that changes in correlated parameters, such as in the TF
relation, should be observable.  As stellar populations age, their
decreasing UV luminosities ought to be reflected through changes in
galactic mass-to-light ratios; conversely, activity through mergers
and rekindled stellar activity leaves similarly observable signatures.

In the future, it may be possible to separate larger samples of
similar galaxies into redshift bins spanning the entire high redshift
range of interest, in order to further quantify the evolution of the
TF intercept.  The DEEP2 survey provides a promising dataset to carry
out such work, and to extend such studies towards examining the role
of individual galaxy variations in the overall scatter of TF and
similar relations.  And while our and other concurrent studies are
capable of revealing evolution of the TF linear fit intercept versus
redshift, in principle, changes in the TF fit slope itself could also
be probed more accurately than has yet been possible -- perhaps
revealing differences in useful observables such as star formation
efficiency versus time.

\section*{Acknowledgments}
We thank the DEEP2 collaboration for making available the high-quality
data gathered over several years at the CFHT and Keck observatories.  We also
appreciate the comments and suggestions of the anonymous referee.  KC acknowledges funding from a PPARC rolling grant, and AJB
acknowledges a Philip Leverhulme Prize.  The analysis pipeline used to
reduce the DEIMOS data was developed at UC Berkeley with support from
NSF grant AST-0071048.

\appendix

\label{lastpage}

\end{document}